\documentclass[useAMS,usenatbib]{mn2e}
\usepackage{graphicx}
\def\apj{ApJ}
\def\apjs{ApJS}

\def\aap{A\&A}

\def\mnras{MNRAS}

\def\pasp{PASP}

\def\rmxaa{Revista Mexicana de Astronomia y Astrofisica}

\title[NIR observations of BCDs]
{Near-infrared spectroscopy of a large sample of low-metallicity blue 
compact dwarf
galaxies}
\author[Y. I. Izotov \& T. X. Thuan]{Y. I.\ Izotov$^{1}$ and
T. X.\ Thuan$^{2}$\\
                $^{1}$Main Astronomical Observatory,
                     Ukrainian National Academy of Sciences,
                     Zabolotnoho 27, Kyiv 03680,  Ukraine\\
                $^{2}$Astronomy Department, University of Virginia, 
                     P.O. Box 400325, Charlottesville, VA 22904-4325\\
}
\begin{document}


\pagerange{\pageref{firstpage}--\pageref{lastpage}} \pubyear{2012}

\maketitle

\label{firstpage}

\begin{abstract}
We present near-infrared (NIR) spectroscopic observations in the wavelength 
range 0.90$\mu$m--2.40$\mu$m of eighteen low-metallicity blue compact dwarf 
(BCD) galaxies and six H {\sc ii} regions in spiral and interacting galaxies. 
Hydrogen and helium emission lines are detected in all spectra, while H$_2$ 
and iron emission lines are detected in most spectra. The NIR data for all 
objects have been supplemented by optical spectra. In all objects, except 
perhaps for the highest metallicity ones, we find that the extinctions $A(V)$ 
in the optical and NIR ranges are similar, implying that the NIR hydrogen 
emission lines in low-metallicity BCDs do not reveal more  star formation 
than seen in the optical. We conclude that emission-line spectra of 
low-metallicity BCDs in the $\sim$0.36--2.40$\mu$m wavelength range are 
emitted by a relatively transparent ionized gas. The  H$_2$ emission line 
fluxes can be accounted for by fluorescence in most of the observed galaxies. 
We find a decrease of the H$_2$ 2.122$\mu$m emission line relative to the 
Br$\gamma$ line with increasing ionization parameter. This indicates an 
efficient destruction of H$_2$ by the stellar UV radiation. The intensities 
of the [Fe {\sc ii}]1.257$\mu$m and 1.644$\mu$m emission lines in the spectra 
of all galaxies, but one, are consistent with the predictions of Cloudy 
stellar photoinization models. There is thus no need to invoke shock 
excitation for these lines, and they are not necessarily shock indicators 
in low-metallicity high-excitation BCDs. The intensity of the 
He {\sc i} 2.058$\mu$m emission line is lower in high-excitation BCDs with 
lower neutral gas column densities and higher turbulent motions.
\end{abstract}


\section{Introduction}\label{sec:INT}

Blue compact dwarf (BCD) galaxies have oxygen abundances that  
vary in the range 12 + log O/H = 7.1 -- 8.3, i.e. 1/45 
--1/3 that of the Sun, if the solar abundance of 
\citet{S15}, 12 + log O/H = 8.76, is adopted. They thus 
constitute excellent nearby laboratories for studying star formation and 
its interaction with the interstellar medium (ISM) in a metal-deficient
environment \citep[see ][for a review]{thuan08}. They are also of cosmological interest because, besides having low metallicities, 
they are also compact, have low-mass (10$^8$-10$^9$ M$_\odot$), high specific star formation rates, high gas masses and clumpy morphologies, making them 
the best local proxies for high-redshift 
Lyman alpha emitting galaxies which also share these properties. 

While BCDs have been extensively studied spectroscopically   
in the optical range, not much recent work has been done in the near-infrared (NIR) 0.9 -- 2.4 $\mu$m range. 
We have started a long-term observational program to study 
the NIR spectroscopic properties of a sample of 
BCDs chosen to cover a large range of metallicities.
The first results of the program, concerning six BCDs, 
have been discussed by \citet*{I09} and \citet{IT11}.   
Here, we present new  $J$, $H$ and $K$ spectroscopic observations 
of eighteen more low-metallicity 
BCDs as well as of H~{\sc ii} regions J0115$-$0051, UM 311, Mrk 1089, Mrk 94, 
J1038+5330, and Mrk 1315 in the spiral or interacting galaxies with slightly 
higher metallicities. 
The  12 + log O/H of the objects in the sample ranges from 7.25 to 8.40. 

\begin{table*}
\caption{General characteristics of galaxies \label{tab1}}
\begin{tabular}{lcclcccl} \hline
Object &R.A. (J2000.0) &Dec. (J2000.0) &$g$, $B$&$M_g$, $M_B$&Redshift&12+logO/H&Comments \\ \hline
HS 0021$+$1347  & 00 24 25.9 & $+$14 04 10& 15.86      &$-$17.81&0.01423& 8.39$^\ddag$&BCD \\
J0115$-$0051    & 01 15 33.8 & $-$00 51 31& 16.51      &$-$14.85&0.00559& 8.37$^\ddag$&H {\sc ii} region in spiral galaxy \\
UM 311          & 01 15 34.4 & $-$00 51 46& 17.90      &$-$13.46&0.00559& 8.33$^\ddag$&H {\sc ii} region in spiral galaxy \\
SHOC 137        & 02 48 15.8 & $-$08 17 24& 16.29      &$-$14.04&0.00469& 7.97$^\ddag$&BCD \\
Mrk 600         & 02 51 04.6 & $+$04 27 14& 15.45$^\dag$&$-$14.73&0.00336& 7.79$^*$   &BCD \\
Mrk 1089        & 05 01 37.7 & $-$04 15 28& 13.40$^\dag$&$-$20.32&0.01341& 8.00$^*$   &interacting galaxies \\
Mrk 94          & 08 37 43.5 & $+$51 38 30& 16.90      &$-$13.50&0.00244& 8.08$^\ddag$&H {\sc ii} region in spiral galaxy \\
I Zw 18         & 09 34 02.0 & $+$55 14 28& 16.08$^\dag$&$-$14.39&0.00251& 7.24$^\ddag$&BCD \\
CGCG 007-025    & 09 44 01.9 & $-$00 38 32& 16.05      &$-$15.91&0.00483& 7.77$^\ddag$&BCD \\
J1038$+$5330    & 10 38 44.8 & $+$53 30 05& 13.15      &$-$17.82&0.00320& 8.30$^\ddag$&H {\sc ii} region in spiral galaxy \\
Haro 3          & 10 45 22.4 & $+$55 57 37& 13.25$^\dag$&$-$17.67&0.00315& ~\,8.35$^{\ddag*}$&BCD \\
Mrk 36          & 11 04 58.3 & $+$29 08 23& 15.70$^\dag$&$-$14.87&0.00216& 7.83$^\ddag$&BCD \\
Mrk 162         & 11 05 08.1 & $+$44 44 47& 14.93      &$-$19.92&0.02154& 8.22$^\ddag$&BCD \\
Mrk 1450        & 11 38 35.7 & $+$57 52 27& 15.29      &$-$15.61&0.00316& 8.11$^*$   &BCD \\
Mrk 193         & 11 55 28.3 & $+$57 39 52& 16.40      &$-$17.94&0.01720& 7.93$^\ddag$&BCD \\
Mrk 1315        & 12 15 18.6 & $+$20 38 27& 16.27      &$-$14.75&0.00283& 8.33$^\ddag$&H {\sc ii} region in spiral galaxy \\
SBS 1222$+$614  & 12 25 05.4 & $+$61 09 11& 14.72      &$-$15.58&0.00236& 8.01$^\ddag$&BCD \\
Mrk 209         & 12 26 15.9 & $+$48 29 37& 15.13      &$-$13.96&0.00094& 7.88$^\ddag$&BCD \\
Mrk 1329        & 12 37 03.0 & $+$06 55 36& 14.40$^\dag$&$-$17.76&0.00544& 8.34$^\ddag$&BCD \\
Mrk 450         & 13 14 48.3 & $+$34 52 51& 14.30$^\dag$&$-$16.59&0.00288& 8.16$^\ddag$&BCD \\
Mrk 259         & 13 28 44.0 & $+$43 55 51& 16.27      &$-$19.13&0.02795& 8.11$^\ddag$&BCD \\
SBS 1415$+$437  & 14 17 01.4 & $+$43 30 05& 15.68$^\dag$&$-$14.42&0.00203& 7.57$^*$   &BCD \\
Mrk 475         & 14 39 05.4 & $+$36 48 22& 15.46$^\dag$&$-$14.57&0.00195& 7.93$^\ddag$&BCD \\
Mrk 487         & 15 37 04.2 & $+$55 15 48& 15.45$^\dag$&$-$14.50&0.00222& 8.06$^\ddag$&BCD \\
\hline
  \end{tabular}

\hbox{
\noindent $^\dag$$B$ magnitude. In all other cases, the $g$ magnitude is given.
}
\hbox{
\noindent $^\ddag$SDSS spectrum is used for oxygen abundance determination.
}
\hbox{
\noindent $^*$3.5m APO spectrum is used for oxygen abundance determination.
}
\hbox{
\noindent $^{\ddag*}$SDSS spectrum of the Haro 3 center and 3.5m APO spectrum of
the offset H {\sc ii} region are used for oxygen abundance
}
\hbox{
\noindent ~~determination. Oxygen abundance is for the galaxy center.
}

  \end{table*}

In this new sample, only a few galaxies have had previous NIR spectral 
observations. However, all these spectra 
covered a more restricted wavelength range, including 
only either the $H$ band and/or the $K$ band, with generally 
a lower signal-to-noise ratio.  
The general characteristics of all 24 objects are given in 
Table \ref{tab1}. With the TripleSpec spectrograph at the 
Apache Point Observatory (APO) 3.5-meter telescope\footnote{The Apache Point 
Observatory 3.5-meter telescope is owned and operated by the Astrophysical 
Research Consortium.}, we can obtain NIR spectra with a simultaneous coverage 
of the $JHK$ bands and significantly more signal-to-noise ratio
than previous spectra. These improved observations will  
allow us to study in more detail the physical conditions in H~{\sc ii} regions.  
The considerably larger sample will also permit us 
to check for the dependence of various physical parameters on metallicity.
We will be able to compare the   
extinctions in the optical and NIR and search for hidden star formation. 
We will also be able to study the excitation mechanisms
of line emission of ionized species in the H {\sc ii} regions, and 
search for molecular hydrogen emission.

We describe the 3.5m APO observations in Section \ref{sec:OBS}. 
In Section \ref{sec:RES}, 
we discuss their extinction, both in the optical and NIR ranges, 
the excitation mechanisms for 
molecular hydrogen and iron emission in the NIR range. We also 
examine how the intensity of the 
He {\sc i} $\lambda$2.058$\mu$m emission depends on 
various physical conditions.  
We summarize our findings in Section \ref{sec:CON}.         

\begin{figure*}
\includegraphics[angle=0,width=0.6\linewidth]{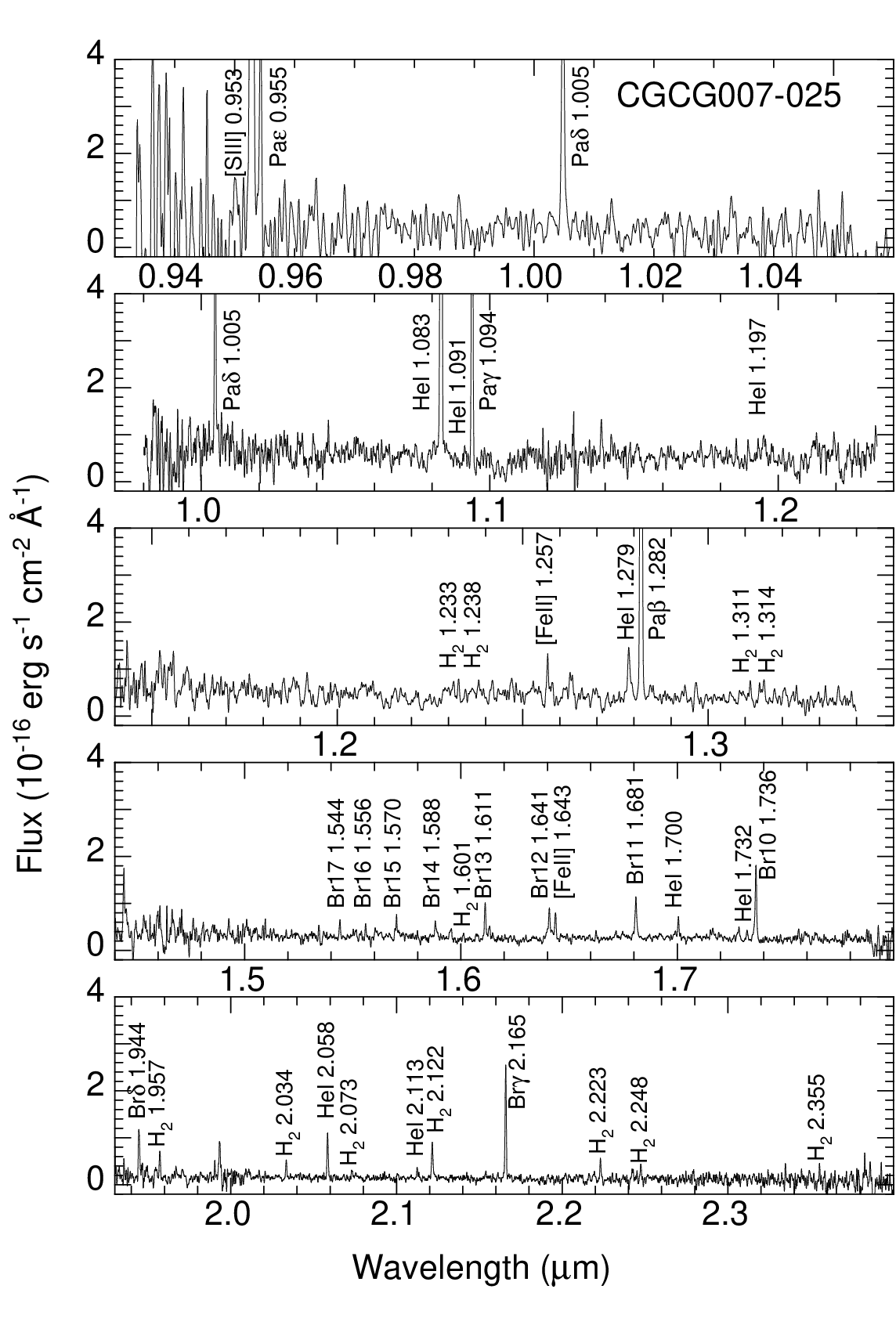}
\caption{As an example of the 3.5m APO/TripleSpec NIR spectra of BCDs,
we show the redshift-corrected spectrum of CGCG007-025 which shows numerous 
H$_2$ emission lines, along with hydrogen, helium and [Fe {\sc ii}] lines.}
\label{fig1}
\end{figure*}

\section{OBSERVATIONS}\label{sec:OBS}

\subsection{Near-infrared observations}\label{sec:NIR}

NIR spectra were obtained with the 3.5 m APO telescope, equipped with the TripleSpec 
spectrograph, on different nights during the 2010-2013 period. 
TripleSpec \citep{W04} is a cross-dispersed NIR 
spectrograph that provides simultaneous continuous wavelength coverage from 
0.90 to 2.46 $\mu$m in five spectral orders during a single exposure. 
A 1\farcs1$\times$43\arcsec\ slit was used, resulting in a resolving power
of 3500. While the present sample contains 24 objects, there are 
two galaxies, Haro 3 and
Mrk 1089, for which spectra were obtained for  
two different star-forming regions within the same object. We therefore analyze a total of 26 NIR spectra.

Several A0V standard stars at airmasses close to those of the objects were 
observed for flux calibration and correction for telluric absorption. 
Spectra of Ar comparison arcs were obtained on the same date 
for wavelength calibration. Since all observed targets
are smaller than the length of the slit, the nod-on-slit 
technique was used to acquire the sky spectrum. Objects were observed
by nodding between two positions A and B along the slit, 
following the ABBA sequence, 
and with an integration time of 200s or 300s at each position. 

We carried out the reduction of the data according to the following procedures.
The two-dimensional spectra were first cleaned for cosmic ray hits 
using the IRAF\footnote{IRAF is 
the Image Reduction and Analysis Facility distributed by the 
National Optical Astronomy Observatory, which is operated by the 
Association of Universities for Research in Astronomy (AURA) under 
cooperative agreement with the National Science Foundation (NSF).} 
routine CRMEDIAN. 
Then all A and B frames were separately coadded and the resulting B frame
was subtracted from the resulting A frame. This procedure is equivalent to 
subtracting each A frame from its time-adjacent B frame and then coadding all 
difference frames. However, some background residuals still remain in the 
coadded spectra because of the sky background short-time variations. Finally, 
the (negative) spectrum at position B was adjusted to the (positive) spectrum 
at position A and subtracted from it. At this stage, the remaining background 
residuals are nearly totally eliminated and the signal-to-noise ratio of the 
final positive spectrum of the object is increased by a factor of $\sqrt[]{2}$ 
because of coadding of the exposures at positions A and B.
The same reduction scheme was applied to the standard stars.

We then use the IRAF
routines IDENTIFY, REIDENTIFY, FITCOORD, TRANSFORM to 
perform wavelength calibration and correction for distortion and tilt for each 
frame. A one-dimensional spectra of all galaxies were extracted from the 
two-dimensional frames using the APALL IRAF routine, within a 
1\farcs1$\times$6\arcsec\ rectangular aperture so as to include the
brightest star-forming regions. 

Flux calibration and correction for telluric absorption were performed 
by first multiplying the 
one-dimensional spectrum of each galaxy by the synthetic absolute 
spectral distributions of standard stars, smoothed to the same
spectral resolution, and then by dividing the result by 
the observed one-dimensional spectrum of the same star.
Since there does not exist any published absolute spectral energy distribution
of standard stars, we have derived them by scaling   
the synthetic absolute spectral energy 
distribution of the star Vega ($\alpha$ Lyrae), of  
similar spectral type A0V, to the brightness of the telluric standard.

As an example, the resulting 
one-dimensional flux-calibrated and redshift-corrected NIR spectrum of the 
BCD CGCG 007-025 is shown in Fig. \ref{fig1}. Strong hydrogen and
helium lines and numerous H$_2$ emission lines are seen.

\begin{table*}
  \caption{Extinction-corrected fluxes of strong hydrogen lines \label{tab2}}
\begin{tabular}{lrrrrrrr}\hline
 &\multicolumn{7}{c}{Galaxy} \\ \cline{2-8}
Line&HS0021$+$1347&J0115$-$0051&UM311&SHOC137&Mrk600&Mrk1089\#1&Mrk1089\#2 \\ \hline
0.410 H$\delta$      &  25.9 &  27.2 &  23.9 &  26.8 &  22.3 &  22.2 &  22.7 \\
0.434 H$\gamma$      &  46.8 &  48.7 &  48.6 &  52.1 &  50.3 &  49.9 &  51.0 \\
0.434 H$\beta$       & 100.0 & 100.0 & 100.0 & 100.0 & 100.0 & 100.0 & 100.0 \\
0.656 H$\alpha$      & 289.6 & 292.9 & 290.1 & 284.9 & 281.7 & 292.3 & 293.4 \\
1.094 P$\gamma$      &   7.6 &   7.6 &   9.1 &   7.8 &  11.4 &   7.9 &   7.2 \\
1.282 P$\beta$       &  16.6 &  12.2 &  18.2 &  15.2 &  20.2 &  16.2 &  15.8 \\
2.165 Br$\gamma$     &   2.9 &   2.1 &   2.1 &   2.8 &   2.9 &   3.1 &   2.4 \\ \hline
Line& Mrk94&IZw18&CGCG007-025&J1038$+$5330&Haro3\#1&Haro3\#2&Mrk36 \\ \hline
0.410 H$\delta$      &  25.7 &  25.2 &  25.6 &  26.3 &  25.9 &  25.0 &  26.6 \\
0.434 H$\gamma$      &  49.0 &  48.4 &  49.3 &  47.7 &  47.3 &  47.6 &  47.4 \\
0.434 H$\beta$       & 100.0 & 100.0 & 100.0 & 100.0 & 100.0 & 100.0 & 100.0 \\
0.656 H$\alpha$      & 287.1 & 279.2 & 284.0 & 296.1 & 291.4 & 293.3 & 281.9 \\
1.094 P$\gamma$      &  13.7 &   6.4 &   8.0 &  10.5 &   8.2 &   9.4 &   9.5 \\
1.282 P$\beta$       &  23.8 &  10.1 &  13.0 &  17.3 &  13.5 &  15.2 &  14.3 \\
2.165 Br$\gamma$     &   3.4 &   1.9 &   2.7 &   2.8 &   2.8 &   3.7 &   1.9 \\ \hline
Line& Mrk162&Mrk1450&Mrk193&Mrk1315&SBS1222$+$614&Mrk209&Mrk1329 \\ \hline
0.410 H$\delta$      &  26.7 &  25.6 &  26.6 &  26.4 &  27.5 &  28.4 &  28.2 \\
0.434 H$\gamma$      &  48.5 &  47.5 &  47.5 &  47.1 &  49.3 &  50.7 &  50.3 \\
0.434 H$\beta$       & 100.0 & 100.0 & 100.0 & 100.0 & 100.0 & 100.0 & 100.0 \\
0.656 H$\alpha$      & 290.6 & 286.7 & 286.1 & 288.7 & 284.9 & 284.2 & 292.7 \\
1.094 P$\gamma$      &   7.8 &   9.7 &   7.0 &  10.3 &   8.1 &   8.6 &   9.8 \\
1.282 P$\beta$       &  12.3 &  13.1 &  13.2 &  15.6 &  14.9 &  14.9 &  14.0 \\
2.165 Br$\gamma$     &   2.8 &   3.2 &   2.8 &   3.9 &   2.6 &   2.5 &   3.1 \\ \hline
Line& Mrk450&Mrk259&SBS1415$+$437&Mrk475&Mrk487
&Case B$^{\rm a}$ \\ \cline{1-7}
0.410 H$\delta$      &  30.4 &  25.7 &  27.3 &  26.2 &  19.6 &  26.2 \\
0.434 H$\gamma$      &  51.7 &  47.5 &  47.7 &  46.2 &  31.7 &  47.3 \\
0.434 H$\beta$       & 100.0 & 100.0 & 100.0 & 100.0 & 100.0 & 100.0 \\
0.656 H$\alpha$      & 291.7 & 288.2 & 283.8 & 277.3 & 283.4 & 279.0 \\
1.094 P$\gamma$      &   9.7 &   9.9 &   6.9 &   6.3 &   6.7 &   8.6 \\
1.282 P$\beta$       &  13.6 &  16.7 &  13.4 &  11.3 &  13.0 &  15.2 \\
2.165 Br$\gamma$     &   2.9 &   3.1 &   2.5 &   1.7 &   2.5 &   2.5 \\ \cline{1-7}\\
\end{tabular}

\hbox{
\hspace{0.5cm}$^{\rm a}$Recombination ratios for $T_e$=15000 K and $N_e$ = 100 cm$^{-3}$ 
from \citet{SH95}.
}
  \end{table*}

\subsection{Optical data}\label{sec:OPT}

To have a more complete physical picture of the extinction and 
of the ionization mechanisms in our galaxies, we have supplemented the 
NIR observations with optical ones. Sloan Digital Sky Survey
(SDSS) spectra are available for all galaxies, except for four 
objects: Mrk600, Mrk 1089, SBS 1415+437, and an offset H~{\sc ii} region 
in Haro~3. Additionally, one object, Mrk 1450, does have a SDSS spectrum, 
but all strong lines are clipped in that spectrum, making it unusable. 
Furthermore, since our galaxies are at low redshifts, 
the [O {\sc ii}] $\lambda$0.373 $\mu$m emission line is not in the observed 
spectral range of their SDSS spectra, except for the SDSS spectrum of Mrk 259. 
This line is essential 
for accurate abundance determinations. Therefore, we have obtained our own 
optical observations for most galaxies with the 3.5 m APO telescope, during 
the course of several nights in the period 2010 -- 2013. This allows us to 
obtain spectra for the 5 objects with no SDSS data, and intensities of 
the [O {\sc ii}] $\lambda$0.373 $\mu$m emission line for most objects. 
 
The 3.5 m APO observations were made with the Dual Imaging 
Spectrograph (DIS) in both the blue and red wavelength ranges. In the blue 
range, we use the B400 grating with a linear dispersion of 
1.83$\times$10$^{-4}$ $\mu$m/pix
and a central wavelength of 0.44 $\mu$m, while in the red range we use the R300 
grating with a linear dispersion 2.31$\times$10$^{-4}$ $\mu$m/pix and a 
central wavelength of 
0.75 $\mu$m. The above instrumental set-up gave a spatial scale along
the slit of 0\farcs4 pixel$^{-1}$, a spectral range $\sim$ 0.36 -- 0.97 $\mu$m 
and a spectral resolution of 7$\times$10$^{-4}$ $\mu$m (FWHM). Several Kitt 
Peak IRS spectroscopic 
standard stars were observed for flux calibration. Spectra of He-Ne-Ar 
comparison arcs were obtained at the beginning or the end of each night for 
wavelength calibration. 
However, we do not have APO observations for CGCG 007-025, UM 311,
Mrk 94, SHOC 137, HS 0021+1347, SBS 1222$+$614, Mrk 162, and
J1038$+$5330. In those cases, we adopted the [O {\sc ii}] $\lambda$0.373$\mu$m 
intensities from \citet{G07} for CGCG 007-025, SHOC 137 and UM 311, 
from \citet{TI05} for Mrk 94, from \citet*{ITL97} for SBS 1222+614, and
from \citet{IT98} for Mrk 162. For the remaining galaxies,
HS 0021+1347 and J1038$+$5330, the [O {\sc ii}] $\lambda$0.373$\mu$m line
intensities were calculated from the [O {\sc ii}] $\lambda$0.720$\mu$m, 
$\lambda$0.730$\mu$m doublet in the SDSS spectra.

The two-dimensional optical spectra were bias subtracted and flat-field 
corrected using IRAF. We then use the IRAF software routines IDENTIFY, 
REIDENTIFY, FITCOORD, TRANSFORM to perform wavelength calibration and 
correction  
for distortion and tilt for each frame. One-dimensional spectra were then 
extracted from each frame using the APALL routine. The sensitivity curve was 
obtained by fitting with a high-order polynomial the observed spectral energy 
distribution of standard stars. Then sensitivity curves obtained
from observations of different stars during the same night were averaged.

With the [O {\sc ii}] $\lambda$0.373$\mu$m line intensity measured from 
the APO/DIS spectra, we have used the SDSS spectra whenever available
and APO spectra otherwise to derive the physical 
conditions and oxygen abundances of all objects, following the procedures 
described by \citet{I06}. The APO spectrum was used for Mrk 1450 because of the 
clipping problem in the SDSS spectrum.  We have preferred to use SDSS instead 
of APO spectra because they are available for 20 out of 24 galaxies 
in our sample. Furthermore, they have a reliable spectrophotometric 
flux calibration, giving  accurate absolute emission-line 
fluxes. On the other hand, only 16 galaxies in our sample possess 
APO/DIS spectra, many of which were obtained in non-photometric conditions, 
introducing uncertainties in the absolute flux calibration. The derived oxygen 
abundances are shown in Table \ref{tab1}. The combination of observations 
obtained with different telescopes and varying spectroscopic apertures may 
introduce uncertainties in those abundances. We estimate these uncertainties 
by comparing the oxygen abundances derived here with those by \citet*{I14}.
They have compiled oxygen abundance determinations for 
many objects in our sample (see their Table 1). It can be seen that, on 
average, the differences between the oxygen abundances in this paper and 
those in Table 1 of \citet{I14} are $\la$ 0.05 dex. This accuracy is quite 
sufficient for our purposes here. On the other hand, extinction coefficients 
do not suffer from this effect since hydrogen line ratios are derived from a 
single spectrum obtained with a single telescope.

We have adjusted the flux levels of the optical and NIR spectra by scaling the fluxes of 
the [S {\sc iii}]$\lambda$0.907$\mu$m emission line in the optical spectrum and 
the [S {\sc iii}]$\lambda$0.953$\mu$m emission line in the NIR spectrum, so that their 
ratio is equal to the theoretical line ratio $\lambda$0.953$\mu$m/$\lambda$0.907$\mu$m 
of 2.486 \citep{A84}. These two lines are close enough in wavelength so that differential 
extinction can be neglected.
This procedure could not be 
applied to two galaxies, Mrk 162 and Mrk 259, the objects with 
the highest redshifts in our sample. In these cases, 
the [S {\sc iii}]$\lambda$0.907$\mu$m emission line is shifted out of 
the SDSS spectral range, so that the adjustment of optical and NIR spectra, 
using the [S {\sc iii}] emission lines, was not possible.
The optical and NIR flux levels of 
the spectra of these two 
galaxies were adjusted by matching their continuum fluxes 
in the overlapping spectral range.

\section{RESULTS AND DISCUSSION}\label{sec:RES}

Emission-line fluxes in both the optical and NIR ranges were measured by using 
the SPLOT routine in IRAF. 
The errors of the line fluxes were calculated from the photon statistics
in the non-flux-calibrated spectra. For most of our galaxies, we have added the errors due to the adjustment of the optical
and NIR spectra to the errors of the NIR lines, using the uncertainties in  
the [S {\sc iii}]$\lambda$0.907$\mu$m and [S {\sc iii}]$\lambda$0.953$\mu$m 
emission-line fluxes. This procedure did not apply to the two objects 
where the continuum adjustment was not done with the [S {\sc iii}] lines, 
Mrk 162 and Mrk 259. Here,    
we have added the uncertainties of the continuum placement
to the errors of the NIR lines.
The emission-line measurements from the adjusted optical and NIR spectra 
are shown in Tables \ref{tab2} -- \ref{tab4}. Table \ref{tab2} gives the extinction-corrected 
fluxes of strong hydrogen lines. Table \ref{tab3} represents the H$_2$ emission-line flux ratios 
and Table \ref{tab4} lists the fluxes of the strongest [Fe {\sc ii}] emission lines. 
All line fluxes are normalized to the H$\beta$ flux.

The presence of many emission lines in the optical and NIR spectra
allows us to study extinction, H$_2$ and [Fe {\sc ii}] emission  
and excitation mechanisms, and He {\sc i} emission 
in galaxies. Combining our present sample of 18 BCDs and 6 H {\sc ii} regions in spiral 
and interacting galaxies with our previous samples containing 6 BCDs \citep{I09,IT11}, we 
have a total sample of 30 objects, large enough to examine various statistical trends.
We discuss these issues in the following sections.

\begin{figure}
\hbox{
\includegraphics[angle=-90,width=0.99\linewidth]{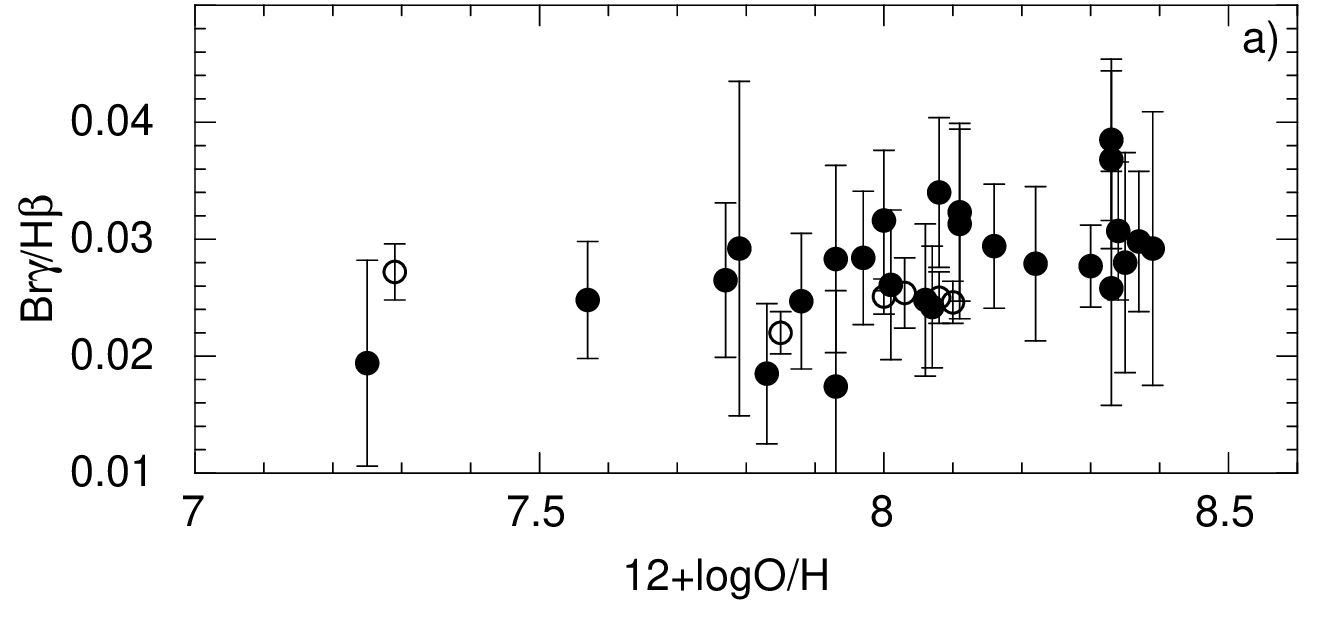}
}
\vspace{0.0cm}
\hbox{
\includegraphics[angle=-90,width=0.99\linewidth]{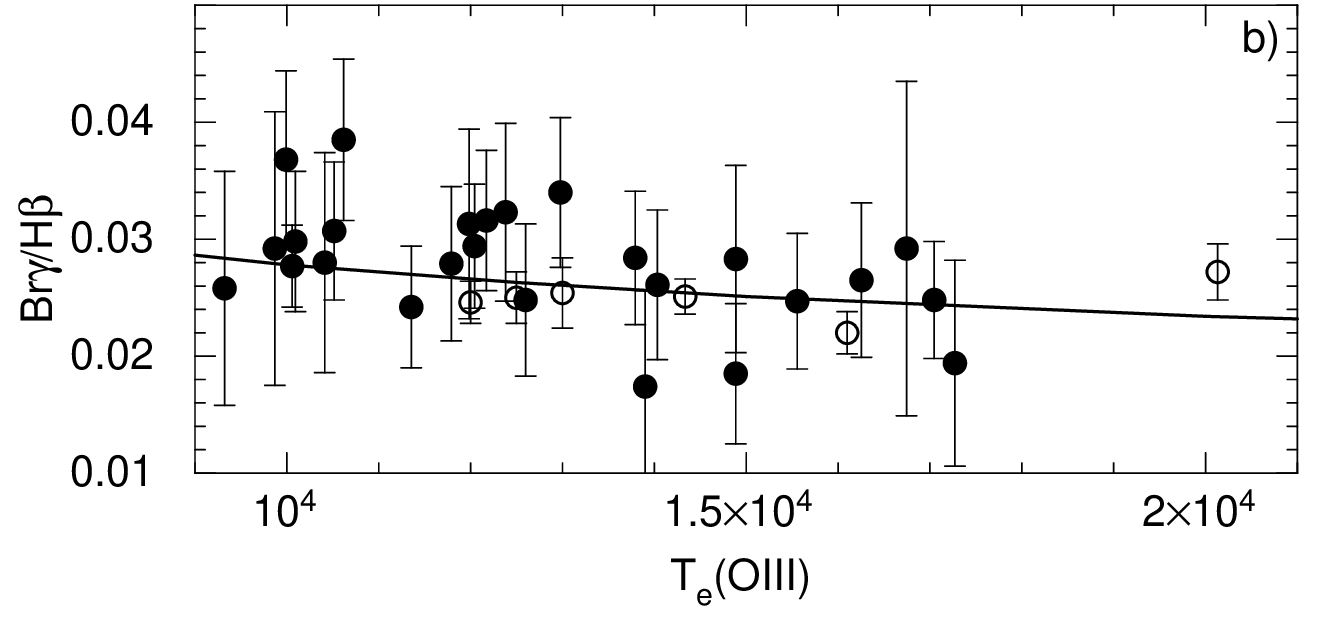}
}
\caption{ The Br$\gamma$/H$\beta$ flux ratio as a function of
(a) the oxygen abundance 12 + log O/H and (b) the electron temperature
$T_{\rm e}$(O {\sc iii}) derived from the 
[O {\sc iii}]$\lambda$0.436/($\lambda$0.496+$\lambda$0.507) flux ratio.
The data from this paper and from our previous work \citep{I09,IT11}
are shown respectively by filled and open circles.
The solid line in (b) shows the theoretical case B recombination ratio 
calculated by \citet{SH95}.}
\label{fig2}
\end{figure}

\begin{table*}
  \caption{H$_2$ emission-line flux ratios relative to the 
H$_2$ $\lambda$2.122$\mu$m flux \label{tab3}}
  \begin{tabular}{lccccccc} \hline
 & \multicolumn{7}{c}{Galaxy} \\ 
\cline{2-8}
Line&HS0021$+$1347&J0115$-$0051&UM311 &SHOC137&Mrk600&Mrk1089\#1&Mrk1089\#1 \\ \hline
1.233 H$_2$ 3-1 S(1)&...&0.4&...&...&...&...&... \\
1.238 H$_2$ 2-0 Q(1)&...&0.6&...&...&...&...&... \\
1.957 H$_2$ 1-0 S(3)&...&0.3&...&...&...&...&... \\
2.034 H$_2$ 1-0 S(2)&...&0.5&0.5&...&...&0.6&0.5 \\
2.073 H$_2$ 2-1 S(3)&...&0.3&...&...&...&0.3&0.4 \\
2.122 H$_2$ 1-0 S(1)&...&1.0&1.0&1.0&1.0&1.0&1.0 \\
2.223 H$_2$ 1-0 S(0)&...&0.4&...&...&...&0.5&0.4 \\
2.248 H$_2$ 2-1 S(1)&...&0.4&...&...&...&0.5&0.4 \\
\hline
Line&Mrk94&IZw18&CGCG007-025&J1038$+$5330&Haro3\#1&Haro3\#2&Mrk36 \\ \hline
1.233 H$_2$ 3-1 S(1)&...&...&0.3&...&...&0.9&... \\
1.238 H$_2$ 2-0 Q(1)&...&...&0.3&...&...&0.7&... \\
1.311 H$_2$ 4-2 S(1)&...&...&0.3&...&...&...&... \\
1.314 H$_2$ 3-1 Q(1)&...&...&0.3&...&...&...&... \\
1.601 H$_2$ 6-4 Q(1)&...&...&0.2&...&...&0.2&... \\
1.957 H$_2$ 1-0 S(3)&...&...&0.6&...&...&...&... \\
2.034 H$_2$ 1-0 S(2)&...&...&0.3&...&...&0.5&... \\
2.073 H$_2$ 2-1 S(3)&...&...&0.1&...&0.4&0.3&... \\
2.122 H$_2$ 1-0 S(1)&...&...&1.0&1.0&1.0&1.0&1.0 \\
2.223 H$_2$ 1-0 S(0)&...&...&0.6&...&0.6&0.3&... \\
2.248 H$_2$ 2-1 S(1)&...&...&0.4&...&0.6&0.5&... \\
2.355 H$_2$ 2-1 S(0)&...&...&0.3&...&...&...&... \\ \hline
Line&Mrk162&Mrk1450&Mrk193&Mrk1315&SBS1222$+$614&Mrk209&Mrk1329 \\ \hline
1.957 H$_2$ 1-0 S(3)&...&...&0.9&...&...&...&... \\
2.034 H$_2$ 1-0 S(2)&0.3&0.7&...&0.5&...&0.8&... \\
2.073 H$_2$ 2-1 S(3)&...&0.5&...&0.7&...&...&... \\
2.122 H$_2$ 1-0 S(1)&1.0&1.0&1.0&1.0&1.0&1.0&1.0 \\
2.223 H$_2$ 1-0 S(0)&0.2&...&...&...&...&...&... \\
2.248 H$_2$ 2-1 S(1)&0.5&...&...&...&...&...&... \\
\hline
Line&Mrk450&Mrk259&SBS1415$+$437&Mrk475&Mrk487&Fluor$^{\rm a}$&Coll$^{\rm a}$ \\ \hline
1.233 H$_2$ 3-1 S(1)&...&...&...&...&0.7&0.5&0.0 \\
1.238 H$_2$ 2-0 Q(1)&...&...&...&...&0.5&0.4&0.0 \\
1.311 H$_2$ 4-2 S(1)&...&...&...&...&...&0.4&0.0 \\
1.314 H$_2$ 3-1 Q(1)&...&...&...&...&...&0.6&0.0 \\
1.601 H$_2$ 6-4 Q(1)&...&...&...&...&...&0.4&0.0 \\
1.957 H$_2$ 1-0 S(3)&...&...&0.9&...&...&...&... \\
2.034 H$_2$ 1-0 S(2)&0.5&...&0.4&...&...&0.5&0.3 \\
2.073 H$_2$ 2-1 S(3)&...&...&0.2&...&...&0.2&0.0 \\
2.122 H$_2$ 1-0 S(1)&1.0&1.0&1.0&1.0&1.0&1.0&1.0 \\
2.223 H$_2$ 1-0 S(0)&...&...&0.3&...&0.5&0.6&0.3 \\
2.248 H$_2$ 2-1 S(1)&...&...&0.4&...&...&0.5&0.0 \\
2.355 H$_2$ 2-1 S(0)&...&...&0.6&...&...&0.3&0.0 \\ \hline
 \end{tabular}

\hbox{
$^{\rm a}$Model values are from \citet{BD87}. We adopt their
model 1 for fluorescent lines and model S1 for collisionally
}
\hbox{
 excited lines.
}
  \end{table*}

\subsection{Optical and NIR extinction and hidden star formation}\label{sec:EXT}

In previous optical-NIR studies of BCDs, except for our own work  
\citep{I09,IT11}, $JHK$ spectra have been observed separately, 
and there has been no wavelength overlap between the 
optical and NIR spectra. This introduces uncertainties, 
due to the use of generally different apertures in the optical and 
NIR observations, and
due to the adjusting of the continuum levels of 
NIR spectra obtained separately in different orders. 
Because our NIR spectra have been obtained simultaneously over the whole 
$JHK$ wavelength range and because it is possible to directly match the 
optical and NIR spectra using [S {\sc iii}] emission line fluxes, these 
adjusting uncertainties are strongly reduced. This allows 
us to directly compare the optical and NIR extinctions and search for any 
hidden extinction at longer wavelengths.  
For comparison, the last column at the bottom 
of Table \ref{tab2} shows the theoretical 
recombination flux ratios, calculated by \citet{SH95} for case B, with
an electron temperature $T_e$ = 15000 K and an electron number density
$N_e$ = 100 cm$^{-3}$. It can be seen that there is general agreement between
the corrected and theoretical recombination values of both optical and 
NIR line fluxes. Therefore the same extinction coefficient 
$C$(H$\beta$) [or $A(V)$] holds over the whole 0.36 -- 2.40 $\mu$m wavelength 
range, implying that there is no appreciable hidden star formation in the 
NIR as compared to in the optical range. 
This appears to be a general result for BCDs \citep[e.g.][]{V00,V02,I09,IT11}. 

Since the amount of dust may be related to the metallicity of the galaxy, one may ask 
whether there is a correlation between the metallicity of the galaxy and its extinction, as 
measured by the observed Br$\gamma$/H$\beta$ flux ratio. In Fig. \ref{fig2}a, we plot the 
Br$\gamma$/H$\beta$ flux ratio against the oxygen abundance 12+logO/H, using the 
present data (filled circles) and that of \citet{I09} and \citet{IT11} (open circles). There is 
a weak trend in the sense of larger Br$\gamma$/H$\beta$ for higher oxygen abundances.
However, the theoretical case B Br$\gamma$/H$\beta$ flux ratio 
does not depend directly on 
the oxygen abundance, but on the electron temperature, which is higher in
low-metallicity galaxies. Therefore we show in Fig. \ref{fig2}b the dependence
of the Br$\gamma$/H$\beta$ flux ratio on the electron temperature 
$T_{\rm e}$(O {\sc iii}), as derived from the 
[O {\sc iii}]$\lambda$0.436/($\lambda$0.496+$\lambda$0.507) flux ratio. 
The solid line indicates the theoretical case B ratio calculated by \citet{SH95}. It is seen
that the Br$\gamma$/H$\beta$ flux ratio is generally consistent with the case B values. 
However, there appers to be a slight systematic excess for galaxies with cool (and hence 
higher-metallicity) H {\sc ii} regions having T$_e$(O~{\sc iii}) $<$13000 K, 
indicating perhaps a small contribution from star-forming regions that are hidden in the 
optical range but seen in the NIR range. But overall, we may conclude that low-metallicity 
BCDs are transparent, in agreement with previous studies by \citet{I09} and \citet{IT11}.

\begin{figure}
\hbox{
\includegraphics[angle=-90,width=0.99\linewidth]{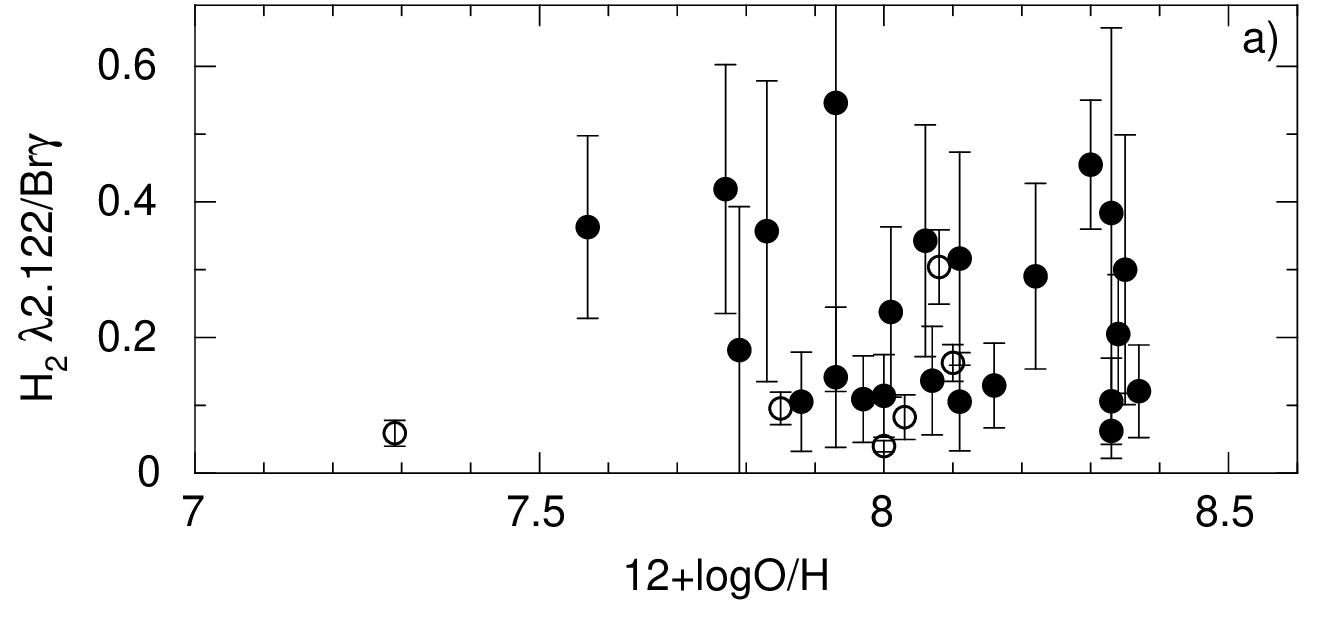}
}
\vspace{-0.1cm}
\hbox{
\includegraphics[angle=-90,width=0.99\linewidth]{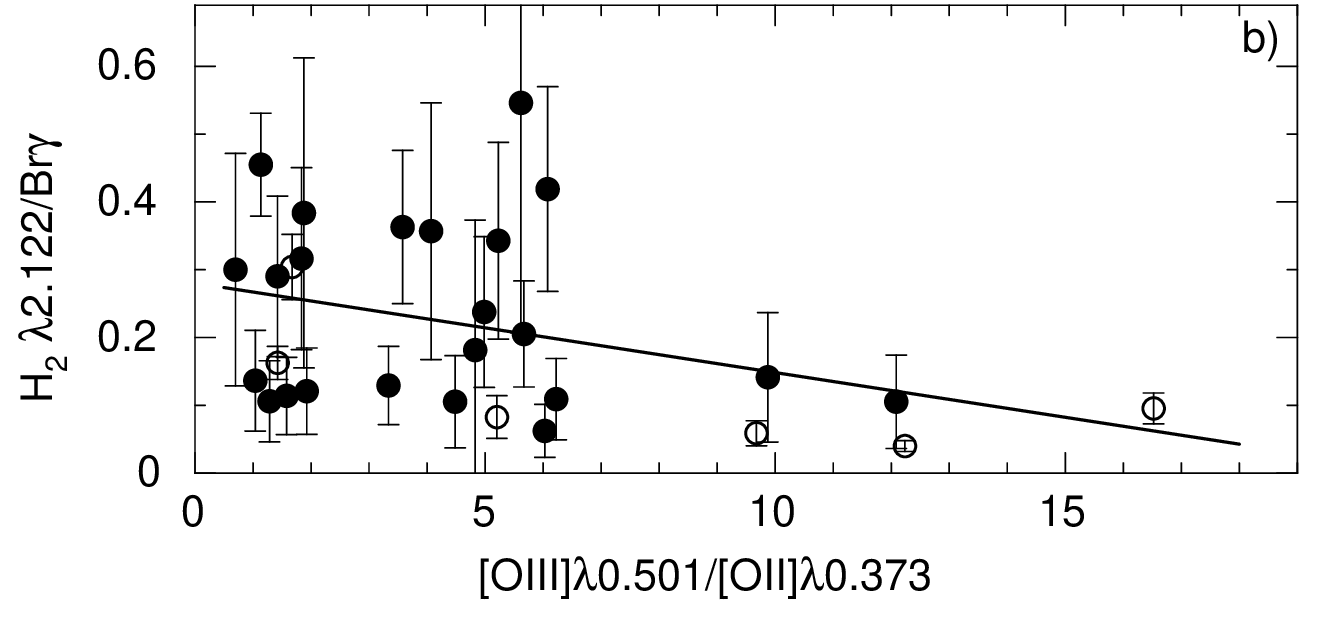}
}
\caption{ Plots of the H$_2$ $\lambda$2.122/Br$\gamma$ flux ratio
as a function of (a) the oxygen abundance 12+logO/H and (b) the 
[O {\sc iii}]$\lambda$0.501/[O {\sc ii}]$\lambda$0.373 flux ratio.
The meaning of the symbols is the same as in Fig. \ref{fig2}. 
The linear regression in (b) is shown by the solid line.}
\label{fig3}
\end{figure}

\begin{table*}
\caption{[Fe {\sc ii}]$\lambda$1.257$\mu$m, [Fe {\sc ii}]$\lambda$1.644$\mu$m
and [Fe {\sc iii}]$\lambda$0.466$\mu$m emission-line fluxes \label{tab4}}
\begin{tabular}{lccc} \hline
Object &100$\times$[Fe {\sc ii}]1.257/H$\beta$ &100$\times$[Fe {\sc ii}]1.644/H$\beta$ &100$\times$[Fe {\sc iii}]0.466/H$\beta$ \\ \hline
HS 0021$+$1347  &     ...     &1.77$\pm$0.88&1.11$\pm$0.71 \\
J0115$-$0051    &0.66$\pm$0.27&0.68$\pm$0.27&0.72$\pm$0.33 \\
UM 311          &1.32$\pm$0.70&1.57$\pm$0.76&0.76$\pm$0.54 \\
SHOC 137        &     ...     &0.41$\pm$0.20&0.43$\pm$0.23 \\
Mrk 1089\#1     &2.25$\pm$0.51&1.43$\pm$0.39&0.99$\pm$0.42 \\
Mrk 1089\#2     &2.47$\pm$0.55&1.62$\pm$0.42&0.96$\pm$0.41 \\
I Zw 18         &0.31$\pm$0.28&     ...     &0.56$\pm$0.39 \\
CGCG 007-025    &0.80$\pm$0.34&0.58$\pm$0.28&0.50$\pm$0.32 \\
J1038$+$5330    &7.69$\pm$0.70&6.62$\pm$0.63&1.41$\pm$0.29 \\
Haro 3\#1       &2.61$\pm$0.91&1.53$\pm$0.67&1.65$\pm$0.79 \\
Haro 3\#2       &1.27$\pm$0.42&1.33$\pm$0.43&1.13$\pm$0.46 \\
Mrk 162         &2.16$\pm$0.58&1.40$\pm$0.45&1.24$\pm$0.50 \\
Mrk 1450        &0.45$\pm$0.26&0.61$\pm$0.31&0.52$\pm$0.31 \\
Mrk 193         &0.89$\pm$0.42&0.71$\pm$0.37&0.55$\pm$0.40 \\
Mrk 1315        &0.18$\pm$0.13&0.35$\pm$0.19&0.17$\pm$0.14 \\
SBS 1222$+$614  &0.99$\pm$0.38&0.54$\pm$0.27&0.72$\pm$0.36 \\
Mrk 450         &0.87$\pm$0.27&0.63$\pm$0.22&0.81$\pm$0.33 \\
Mrk 487         &0.75$\pm$0.34&     ...     &0.55$\pm$0.29 \\ \hline
  \end{tabular}
  \end{table*}

\subsection{H$_2$ emission}\label{sec:H2}

Molecular hydrogen lines do not originate in the H {\sc ii} region, but in 
neutral molecular clouds. In the NIR, they are excited by two
mechanisms. The first one is the thermal mechanism consisting of 
collisions between neutral species (e.g., H, H$_2$), 
resulting from large-scale shocks 
driven by powerful stellar winds, supernova remnants 
or molecular cloud collisions. The second one is 
the fluorescent mechanism due to absorption of ultraviolet photons.
It is known that these two mechanisms excite mostly different
roto-vibrational levels of H$_2$. By comparing the observed line ratios with 
those predicted by models such as those calculated by \citet{BD87}, it is 
possible to discriminate between the two processes. 
In particular, line emission from the vibrational level $v$=2 and higher 
vibrational levels are virtually absent in collisionally excited spectra, 
while they are relatively strong in fluorescent spectra.

H$_2$ emission lines are detected in 23 NIR spectra of 21 objects 
(Mrk 1089 and Haro 3 have two spectra each, Table \ref{tab3}),
but they are not seen in the spectra of HS0021$+$1347, I~Zw~18 and Mrk 94,
possibly because of the weakness of these lines and spectra with insufficient
signal-to-noise ratio. This corresponds to a H$_2$ detection rate of 88\% for our 
sample. We note that \citet{H10} also detected with mid-infrared Spitzer/IRS observations 
rotational transitions of H$_2$ in the warm neutral phase of the ISM of more than a third 
of their sample of 22 BCDs. Although this H$_2$ detection rate is lower than ours because 
of the lower signal-to-noise ratio of the IRS spectra, the Spitzer data also show that 
H$_2$ molecules are present in the warm neutral phase of the ISM of BCDs.

Four objects in our total BCD sample -- I Zw 18, Mrk~209$\equiv$I Zw 36, Mrk 59 and 
SBS~0335--052E -- have also been observed by the Far Ultraviolet Spectroscopic Explorer 
(FUSE) satellite in an attempt to detect H$_2$ in absorption spectra. However, none 
showed H$_2$ absorption lines \citep*[e.g.][]{TL05}. Yet, except for I Zw 18, the other three 
BCDs all show H$_2$ NIR emission lines 
\citep[Mrk 59, SBS~0335--052E,][ and Mrk 209, this paper]{I09,IT11}. 
The absence of  H$_2$ absorption lines in FUSE spectra implies that 
the warm H$_2$ detected through IR emission must be quite clumpy. The FUSE 
observations are not sensitive to such a clumpy H$_2$ distribution as they can only probe 
the transparent UV sight lines, and are not able to penetrate dense clouds with 
$N$(H$_2$) $\ga$ 10$^{20}$ cm$^{-2}$ \citep{H04}.
    
We next discuss the H$_2$ excitation mechanism in our objects.
For eight objects, SHOC 137, Mrk 600, J1038$+$5330, Mrk 36, SBS 1222$+$614, 
Mrk 1329, Mrk 259, and Mrk 475, only one H$_2$ emission line, the 2.122 $\mu$m 
1-0 S(1) line, is definitely detected. As a consequence, no conclusion about the H$_2$ 
excitation mechanism can be drawn for these galaxies. 
On the other hand, several H$_2$ emission lines are detected in 
the remaining 15 spectra of 13 objects. Comparison of the observed and theoretical ratios 
calculated by \citet{BD87}, 
in the cases of fluorescent and collisional excitation (the last two columns at the bottom of 
Table~\ref{tab3}), indicate that the observed line flux ratios are
in general agreement with those predicted for fluorescent excitation. Our finding is 
in agreement with the conclusions of \citet{V00} for SBS 0335--052E,
of \citet{V08} for II Zw 40 and of 
\citet{I09} and \citet{IT11} for a sample of six BCDs. Those authors also found
that the fluorescence process is the main excitation mechanism of H$_2$ lines. 
This conclusion appears to hold for BCDs with high-excitation spectra.

Does the H$_2$ emission depend on some global parameter of the galaxy? 
The flux of the strongest H$_2$ $\lambda$2.122$\mu$m emission line 
does not depend on oxygen abundance (Fig.~\ref{fig3}a). However, since
H$_2$ is a fragile species and can be easily destroyed by stellar
UV radiation, we may expect the H$_2$ flux to depend on the
ionization parameter. A good observational measure of the ionization parameter is the 
ratio of oxygen lines [O~{\sc iii}]$\lambda$0.501/[O~{\sc ii}]$\lambda$0.373. 
We plot in Fig.~\ref{fig3}b the H$_2$ $\lambda$2.122$\mu$m/Br$\gamma$ flux ratio 
against the [O~{\sc iii}]$\lambda$0.501/[O~{\sc ii}]$\lambda$0.373 ratio. 
There is a weak
correlation between the two quantities in the sense that the 
H$_2$ $\lambda$2.122$\mu$m/Br$\gamma$ flux ratio decreases with increasing 
[O~{\sc iii}]$\lambda$0.501/[O~{\sc ii}]$\lambda$0.373 ratios. The linear regression is 
shown by the solid line. There appears to be a floor at H$_2$/Br$\gamma$ 
$\sim$ 0.05 -- 0.1 with a weak statistical trend for lower H$_2$ 2.12$\mu$m/Br$\gamma$ with 
increasing ionization parameter, suggesting an efficient destruction of 
H$_2$ molecules by the UV radiation in high-excitation H {\sc ii} regions.

\begin{figure}
\hbox{
\includegraphics[angle=-90,width=0.99\linewidth]{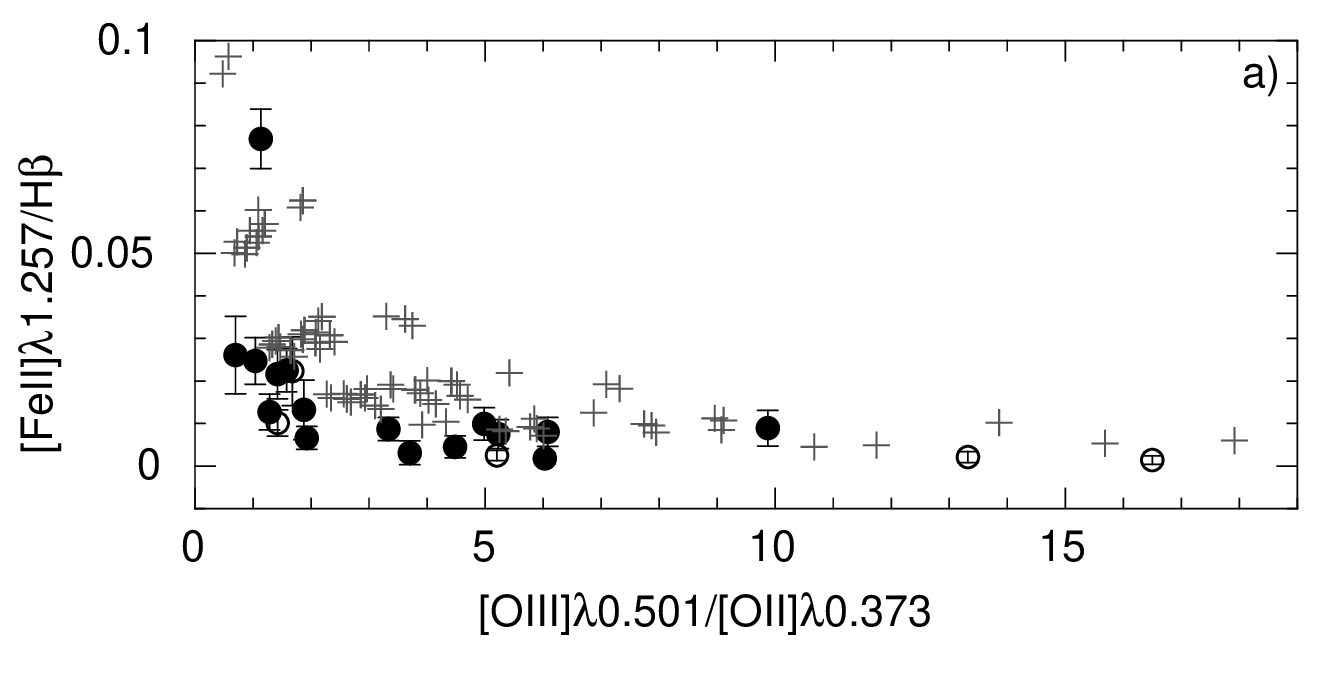}
}
\vspace{-0.1cm}
\hbox{
\includegraphics[angle=-90,width=0.99\linewidth]{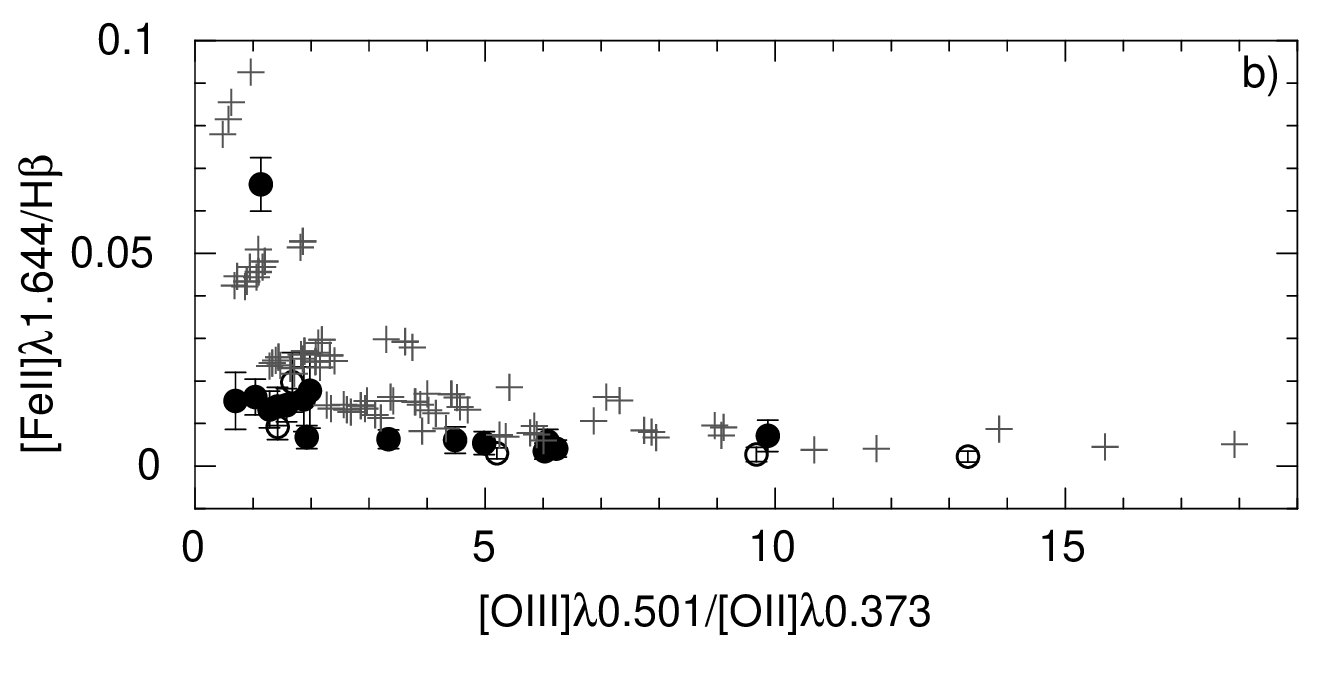}
}
\caption{Dependence on the 
[O {\sc iii}]$\lambda$0.501/[O {\sc ii}]$\lambda$0.373 flux ratio
of the: (a) [Fe {\sc ii}]$\lambda$1.257/H$\beta$ flux ratio; and
(b) [Fe {\sc ii}]$\lambda$1.644/H$\beta$ flux ratio. The meaning of 
filled and open circles is the same as in Fig. \ref{fig2}. Theoretical
flux ratios calculated with the Cloudy code 
\citep[version c13.01, ][]{F98,F13} for oxygen 
abundances 12+logO/H in the 7.6 -- 8.3 range, and for 
various ionization 
parameters, are shown by crosses.}
\label{fig4}
\end{figure}

\begin{figure}
\hbox{
\includegraphics[angle=-90,width=0.99\linewidth]{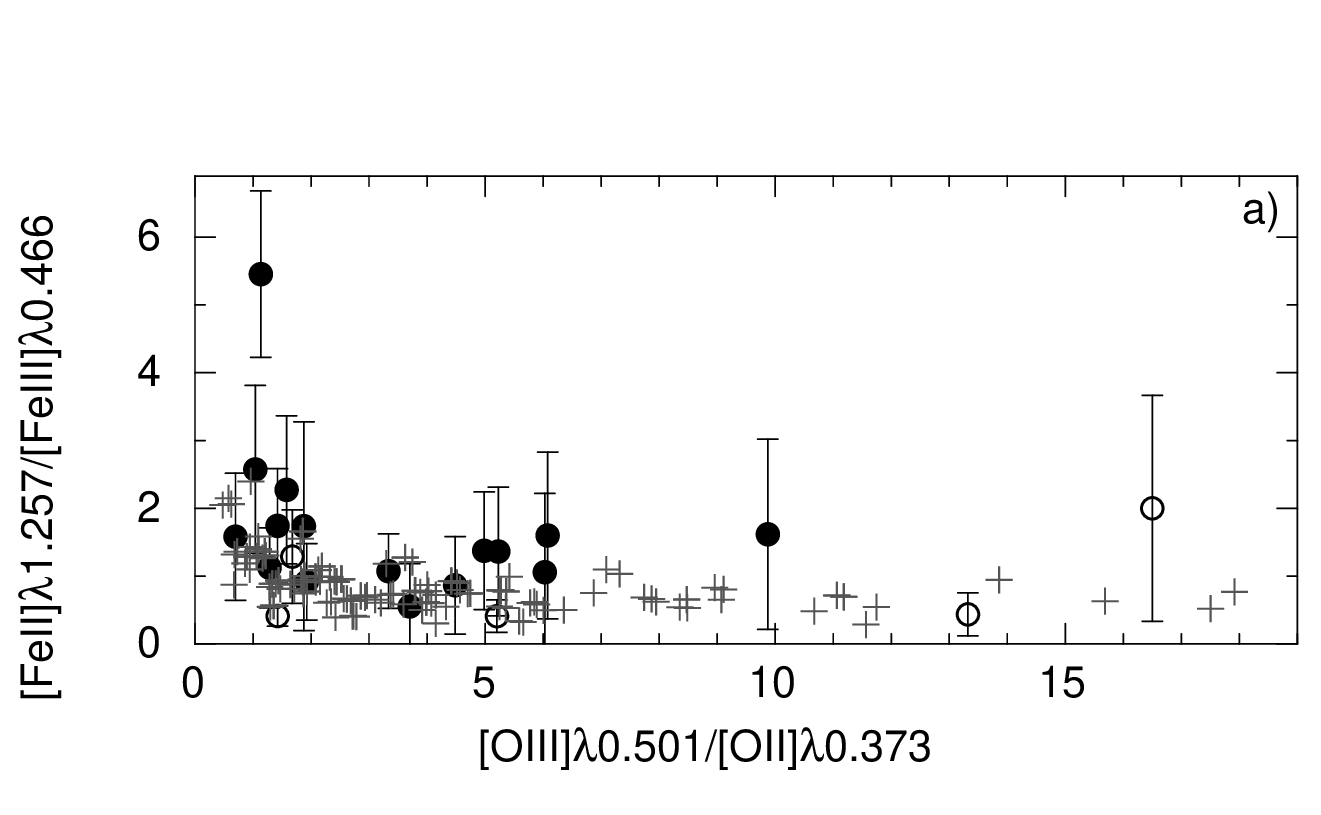}
}
\vspace{-1.0cm}
\hbox{
\includegraphics[angle=-90,width=0.99\linewidth]{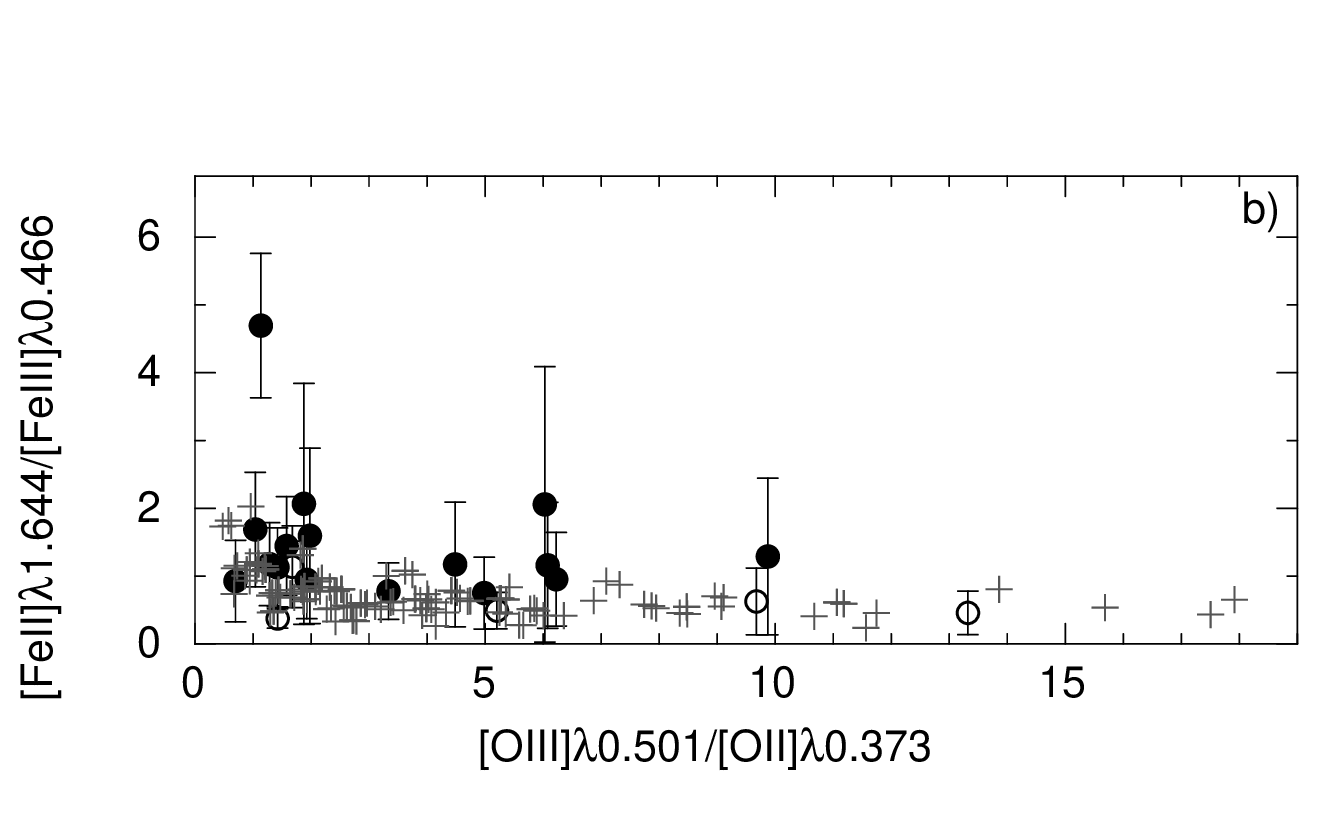}
}
\caption{The dependence on the 
[O~{\sc iii}]$\lambda$0.501/[O~{\sc ii}]$\lambda$0.373 flux ratio
of the : (a) [Fe~{\sc ii}]$\lambda$1.257/[Fe~{\sc iii}]$\lambda$0.466
flux ratio and
(b) [Fe~{\sc ii}]$\lambda$1.644/[Fe~{\sc iii}]$\lambda$0.466 
flux ratio. The meaning of 
filled and open circles is the same as in Fig. \ref{fig2}. Theoretical
flux ratios calculated with the Cloudy code 
\citep[version c13.01, ][]{F98,F13} for oxygen 
abundances 12+logO/H in the 7.6 -- 8.3 range, and for various ionization 
parameters, are shown by crosses.}
\label{fig5}
\end{figure}

\subsection{[Fe {\sc ii}] line emission}\label{sec:FeII}

The [Fe {\sc ii}] $\lambda$1.257 and $\lambda$1.644 $\mu$m emission
lines have often been used to estimate the relative importance of shock excitation and 
photoionization in galaxies \citep*{MO88,O90,O01}. 
We have detected either one of the lines or both in 18 spectra of our galaxy sample 
(Table \ref{tab4}). The fluxes of these lines are considerably 
smaller in regions where photoionization dominates ($\sim$ 1\% of the H$\beta$
flux) than in those where shock excitation plays the main role, such as in SN remnants. 
Shock models by \citet{A08} with shocks propagating in the neutral gas predict  
the [Fe {\sc ii}] $\lambda$1.257 and $\lambda$1.644 $\mu$m emission line fluxes
to be $\sim$ 10\% -- 60\% that of H$\beta$.

Figs. \ref{fig4}ab show that the [Fe {\sc ii}] $\lambda$1.257 $\mu$m and 
$\lambda$1.644 $\mu$m to H$\beta$ line ratios have the very 
low values of $\la$ 0.01 in high-excitation H {\sc ii} regions,
with [O {\sc iii}]$\lambda$0.501/[O {\sc ii}]$\lambda$0.373 $\ga$ 3. 
However, in lower-excitation H {\sc ii}
regions, the [Fe {\sc ii}]/H$\beta$ ratios are 
considerably higher and they steeply increase for
[O {\sc iii}]$\lambda$0.501/[O {\sc ii}]$\lambda$0.373 $\la$ 1. 
Does this mean that SN shock excitation plays an important role
in galaxies with low-excitation H {\sc ii} regions? 

Comparison of the observed flux ratios (filled and open circles) with the theoretical values 
from the stellar photoionization Cloudy models (crosses) in Fig. \ref{fig4} shows that, even 
without any contribution of shock excitation from SNe,
photoionization models overpredict the [Fe {\sc ii}] $\lambda$1.257 and 
$\lambda$1.644 $\mu$m line fluxes by factors of $\sim$1.5 -- 3. 
An exception is the H~{\sc ii} region J1038$+$5330 in the spiral galaxy 
NGC 3310, 
the most outlying object in Figs. \ref{fig4}ab, 
where the extinction-corrected [Fe {\sc ii}] line fluxes reach the high values of 
$\sim$ 7\% -- 8\% that of H$\beta$.
However, even these high values are still in agreement with 
the predictions of some stellar photoionization H {\sc ii} region models.
     
The relations in Fig. \ref{fig4} may be affected by variations of iron
abundances relative to oxygen abundances. However, the emission-line
flux ratios of iron in different stages of ionization should be free
of this effect. Furthermore, these ratios should also be less sensitive to
iron depletion onto dust. We therefore show in Fig. \ref{fig5} the 
[Fe {\sc ii}]$\lambda$1.257/[Fe {\sc iii}]$\lambda$0.466 and
[Fe {\sc ii}]$\lambda$1.644/[Fe {\sc iii}]$\lambda$0.466 flux ratios
as a function of the [O {\sc iii}]$\lambda$0.501/[O {\sc ii}]$\lambda$0.373 
flux ratio. The latter is a measure, as discussed previously, of the ionization parameter. 
We also compare the observed ratios with theoretical predictions from Cloudy models.
At variance with Fig. \ref{fig4} in which the observed values are less than the predicted 
values, the observed values in Fig. \ref{fig5} are in excess of model predictions. 
However, within the errors, observed and modelled flux ratios are still consistent.  
The only exception is again the outlier H~{\sc ii} region J1038$+$5330. The 
[Fe {\sc ii}]/[Fe {\sc iii}] flux ratios
in this object cannot be reproduced by any stellar photoionization H {\sc ii}
region model. Apparently, shocks are important contributors to the ionization
and excitation of iron lines in this H~{\sc ii} region. 

\citet{A08} do not give [Fe~{\sc iii}]$\lambda$0.466 fluxes for their shock 
models, which prevents us from  estimating the contribution of shocks in J1038$+$5330.
For the remaining galaxies it appears that there is no need to
invoke any shock ionizing radiation. We conclude that SN 
shock excitation does not generally play an important role for the 
excitation and ionization of iron lines in BCDs. In these objects, the 
presence of [Fe~{\sc ii}] lines does not necessarily imply the presence of 
shocks.

Some caveats regarding the model predictions should be mentioned.
\citet{A08} considered only shocks propagating in a neutral ISM. The
situation can be quite different in our galaxies, where shocks may propagate
in an ionized medium. The most sensitive ion to these differences is 
[Fe {\sc ii}] and therefore the available shock model predictions for [Fe {\sc ii}] emission 
may not be valid for our objects. Finally, at the present time, 
there is no H {\sc ii} region model that self-consistently treats together the effects of
stellar and shock ionizing radiation. All our comparisons are based only on the 
superposition of stellar photoionized H~{\sc ii} region models and shock models, 
calculated separately, a procedure which may not be appropriate.

\subsection{He {\sc i} $\lambda$2.058$\mu$m emission}\label{sec:HeI}

The He {\sc i} $\lambda$2.058$\mu$m line is one of the strongest emission lines 
in the $K$-band spectra of star-forming galaxies. It is observed in all objects
in our sample. Since this line is emitted in the 2$^1P$$\rightarrow$2$^1S$ transition,
its intensity depends on the conditions in which the permitted resonance 
He {\sc i} Ly$\alpha$ 2$^1P$$\rightarrow$1$^1S$ $\lambda$584\AA\ line is scattered 
\citep{F99}.

The He {\sc i} Ly$\alpha$ $\lambda$584\AA\ photon can be converted into a
$\lambda$2.058 $\mu$m photon with a probability
of 10$^{-3}$ per scattering \citep{F80}. An atom in 2$^1S$ state can either
undergo a two-photon transition to the ground state 1$^1S$ or re-emit 
a $\lambda$584\AA\ photon after a 2$^1S$$\rightarrow$2$^1P$ collisional 
transition. Therefore, the intensity of the He {\sc i} $\lambda$2.058 $\mu$m 
line decreases with increasing dust-to-gas mass ratio due to a more efficient 
destruction of the $\lambda$584\AA\ photons. It can also decrease with 
decreasing neutral gas column density and increasing turbulent velocity 
because both make the $\lambda$584\AA\ photons escape the H {\sc ii} region 
more easily \citep{F99}. Thus, the intensity of the 
He {\sc i} $\lambda$2.058 $\mu$m line can potentially be a good diagnostic 
for resonant line scattering in H~{\sc ii} regions.

\begin{figure}
\hbox{
\includegraphics[angle=-90,width=0.99\linewidth]{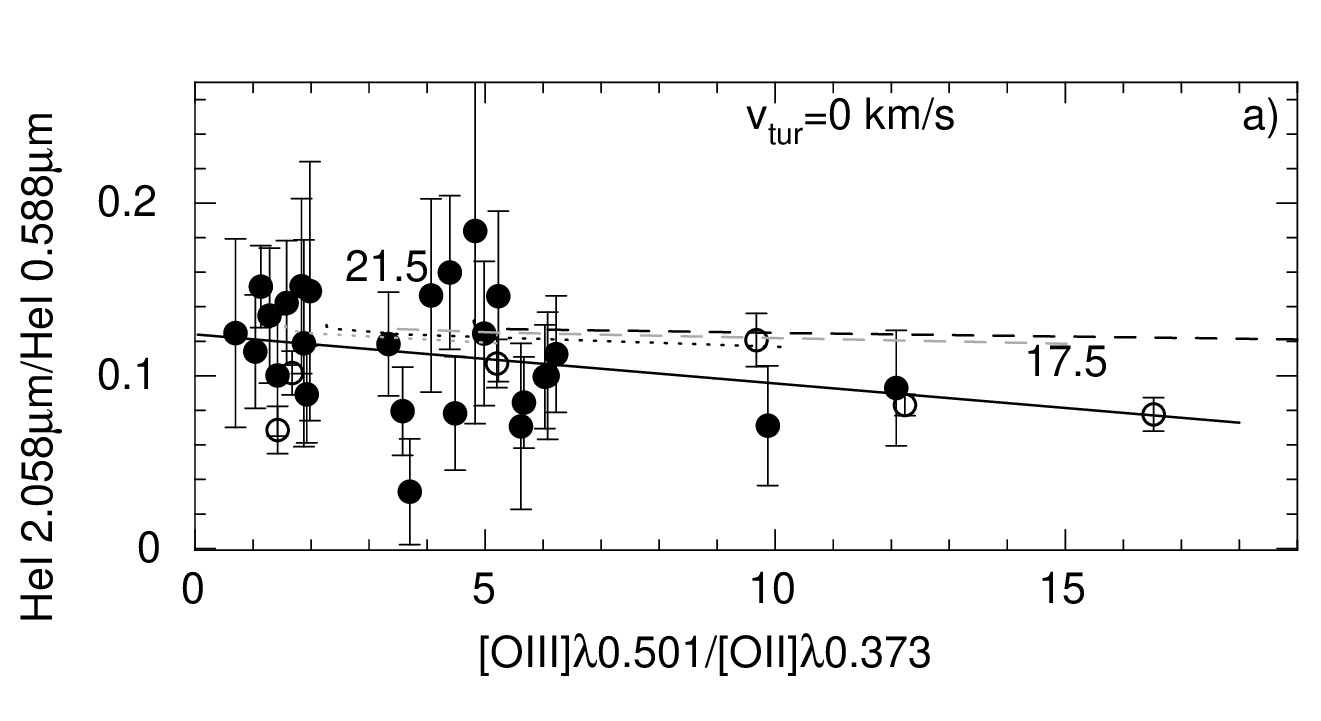}
}
\vspace{-0.3cm}
\hbox{
\includegraphics[angle=-90,width=0.99\linewidth]{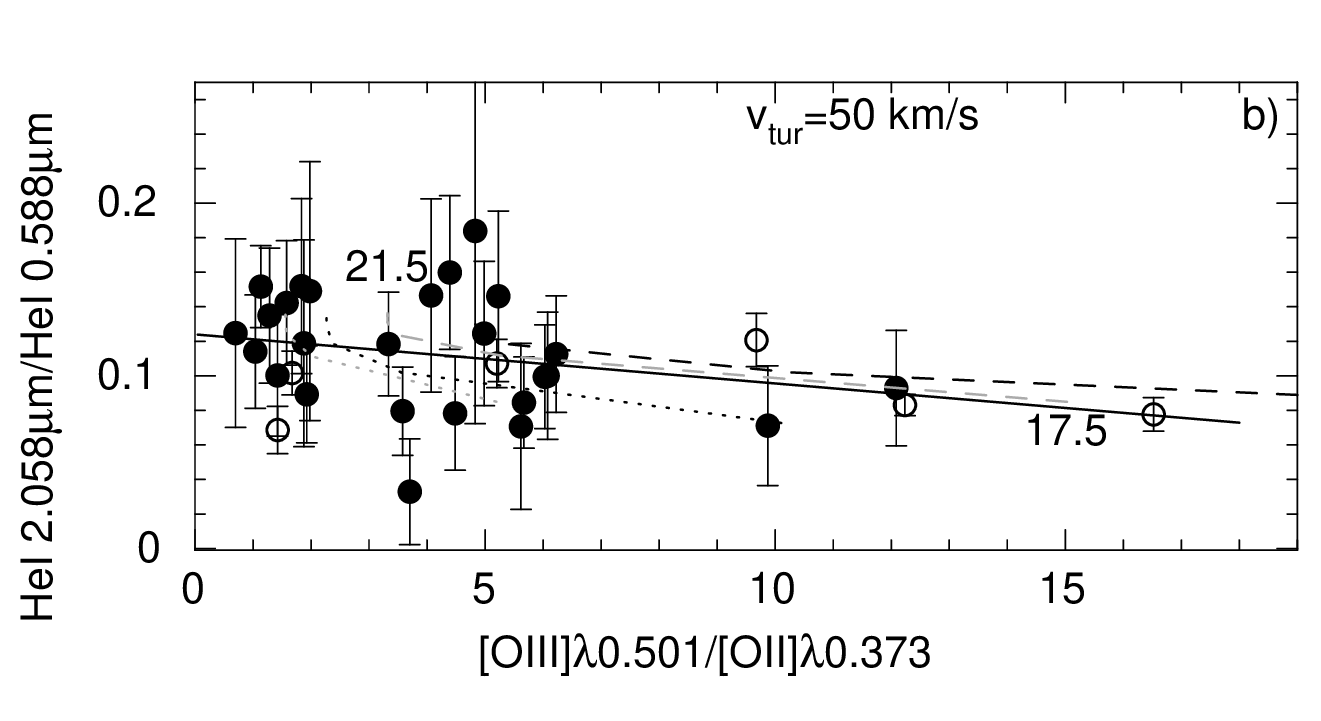}
}
\vspace{-0.3cm}
\hbox{
\includegraphics[angle=-90,width=0.99\linewidth]{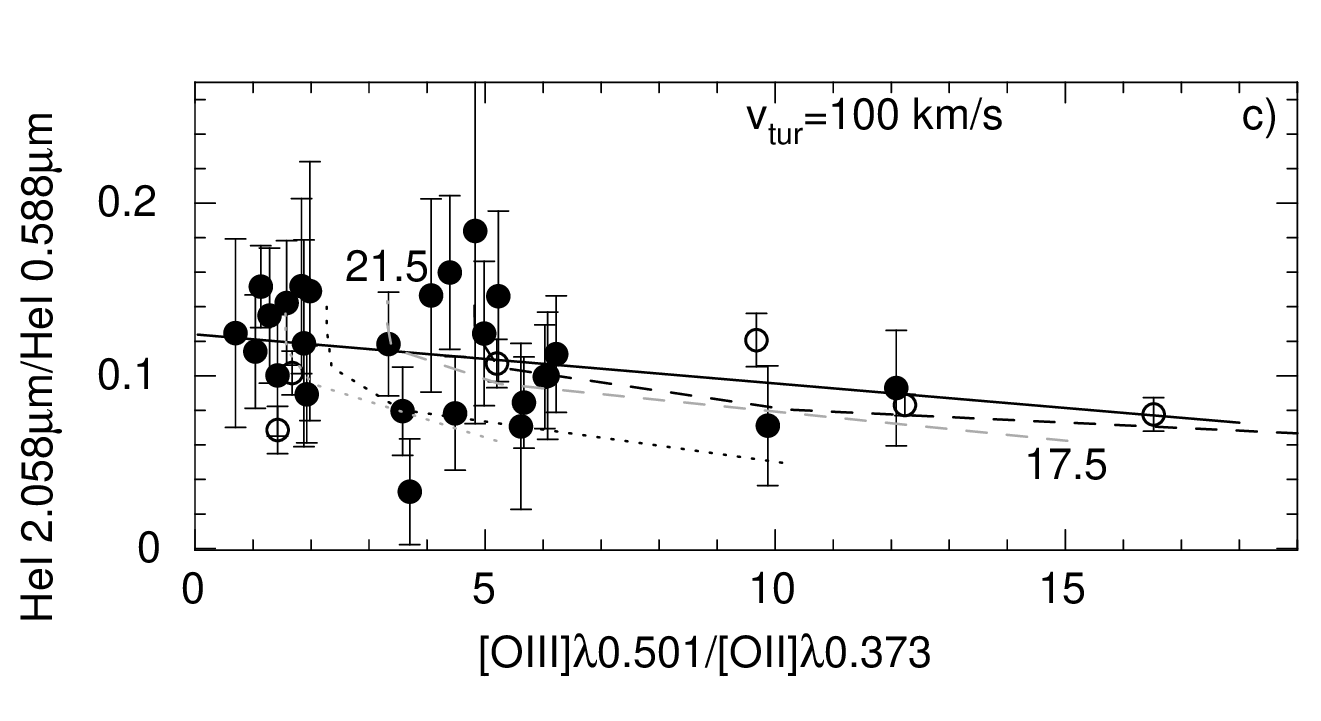}
}
\caption{a) The He {\sc i} 2.058/He {\sc i} 0.588 flux ratio as a 
function of the 
[O {\sc iii}]/[O {\sc ii}] flux ratio. The black solid line is the maximum-likelihood fit  
to the data. The black dashed and dotted lines show He {\sc i} 2.058/He {\sc i} 0.588 
flux ratios calculated with the Cloudy code for a number of ionising photons 
equal to 10$^{53}$ s$^{-1}$ (dashed line) and 10$^{52}$ s$^{-1}$ (dotted line), and 
for various values of the H {\sc i} column density,
varying in log scale from 17.5 to 21.5, with
increments of 0.5 dex. The turbulent velocity
is set to 0 km s$^{-1}$, while a starburst age of 2 Myr is adopted. 
Grey dashed and dotted lines are the same as black dashed and dotted 
lines, but for starburst age of 3 Myr. 
b) Same as a), but Cloudy models are calculated with a turbulent velocity
of 50 km s$^{-1}$.
c) Same as a), but Cloudy models are calculated with a turbulent velocity
of 100 km s$^{-1}$.
}
\label{fig6}
\end{figure}

It is seen in Fig. \ref{fig6}a that the
He {\sc i} 2.058$\mu$m/He {\sc i} 0.588$\mu$m flux ratio decreases with increasing
[O {\sc iii}] 0.501$\mu$m/[O {\sc ii}] 0.373$\mu$m flux ratio,
an indicator of the ionization parameter. This dependence can be 
approximated by the linear maximum-likelihood relation shown by the solid black line. 
For comparison, we show different model dependences calculated with the
Cloudy code for various neutral hydrogen column densities in the range $N$(H {\sc i}) =
10$^{17.5}$ -- 10$^{21.5}$ cm$^{-2}$ [$N$(H {\sc i}) increases from right to left], 
a turbulent velocity of 0 km s$^{-1}$, two values of the starburst
age, 2 Myr (black dashed and dotted lines), 3 Myr (grey lines),
and two values of the number of ionizing photons $Q_{\rm H}$ = 10$^{52}$ s$^{-1}$ 
(dotted lines) and 10$^{53}$ s$^{-1}$ (dashed lines). The oxygen abundance 
12+logO/H = 8.0 is adopted for all models.

This set of Cloudy models with no turbulent velocity predicts a constant 
He {\sc i} 2.058$\mu$m/He {\sc i} 0.588$\mu$m flux ratio, independently of
the [O {\sc iii}] 0.501$\mu$m/[O {\sc ii}] 0.373$\mu$m flux ratio and the neutral 
hydrogen column density, and hence does not reproduce the observed trend.
This is because the optical depth of the resonance transition 
2$^1S$$\rightarrow$2$^1P$ in these models is always sufficiently 
high ($>$10$^{3}$), even in models with the lowest 
neutral hydrogen column density [$N$(H {\sc i}) = 10$^{17.5}$ cm$^{-2}$], 
so that production of the He {\sc i} 2.058$\mu$m photons is always efficient.

On the other hand, models with the more realistic turbulent velocities of 50 km s$^{-1}$ and
100 km s$^{-1}$, values that are observed in extragalactic H {\sc ii} regions \citep{Ch14}, 
do much better in reproducing the observed relation (Figs. \ref{fig6}bc). 
We thus conclude that the observed trend of the
He {\sc i} 2.058$\mu$m/He {\sc i} 0.588$\mu$m flux ratio 
with the [O~{\sc iii}] 0.501$\mu$m/[O~{\sc ii}] 0.373$\mu$m flux ratio can be understood 
as due to the combined effects of turbulent motions in H {\sc ii} regions and 
a finite neutral hydrogen column density. The latter likely decreases with increasing
[O~{\sc iii}] 0.501$\mu$m/[O~{\sc ii}] 0.373$\mu$m flux ratios, 
indicating a higher ionization parameter.

\section{SUMMARY AND CONCLUSIONS}\label{sec:CON}

We present here near-infrared (NIR) spectroscopic observations of eighteen blue compact dwarf (BCD) galaxies and six H {\sc ii} regions
in spiral and interacting galaxies obtained with the 3.5m APO telescope, equipped with the TripleSpec 
spectrograph. Two galaxies have each two different star-forming regions observed, bringing 
the total number of NIR spectra to twenty six. The NIR data have been supplemented
by Sloan Digital Sky Survey (SDSS) spectra and some spectra obtained with the 
same 3.5m APO telescope and the optical DIS spectrograph.
For most galaxies, the 
flux levels of the optical and NIR spectra are adjusted by scaling the fluxes of the 
[S {\sc iii}]$\lambda$0.907$\mu$m emission line in the optical spectrum and the 
[S {\sc iii}]$\lambda$0.953$\mu$m emission line in the NIR spectrum, so that their ratio 
is equal to the theoretical value. Only for two galaxies (those with the highest redshifts) 
were continua used to match the optical and NIR spectra. To increase the statistics, we 
have combined the present observations with our previously published NIR data on six 
objects, resulting in a total sample of 30  objects.

We have arrived at the following conclusions:

1. In all galaxies, except perhaps for the highest-metallicity ones, 
the extinction $A$($V$) derived in the optical 
is the same as the one derived in the NIR for the ionized gas 
regions traced by emission lines. 
The NIR emission lines of low-metallicity BCDs 
do not probe more extinct regions with 
hidden star formation as compared to the optical emission lines.
For the highest-metallicity  H~{\sc ii} regions, there is a suggestion that 
the extinction in the NIR wavelength range corresponds to a slightly 
higher $A$($V$) than in the optical range.

2. We have detected molecular hydrogen emission lines in twenty three 
NIR spectra, i.e. in 88\% of the objects in the present sample.
 Comparison of the observed fluxes with modelled ones 
suggests that the main excitation mechanism of H$_2$ emission
in all objects is fluorescence. We find that the flux of the strongest
H$_2$ emission line decreases when the ionization parameter increases,
implying that the UV ionizing radiation of H {\sc ii} regions is efficient 
at destroying molecular hydrogen.

3. A Cloudy model with a pure stellar ionizing radiation
reproduces well the observed fluxes of emission
lines in both the optical and NIR ranges for all galaxies, but one. 
Shocks are likely present in the 
H {\sc ii} regions, but they play a minor role in the ionization. This means 
that the [Fe {\sc ii}] $\lambda$1.257 and $\lambda$1.644 $\mu$m emission lines, 
often used as SN shock indicators in low-excitation high-metallicity starburst 
galaxies, cannot play such a role in high-excitation low-metallicity 
H~{\sc ii} regions. However, we find that shocks could be important 
contributors to the emission of [Fe {\sc ii}] lines in the brightest H~{\sc ii} region J1038$+$5330 of the spiral galaxy NGC 3310.

4. We find that the intensity of the He {\sc i} $\lambda$2.058$\mu$m emission
line is lower in high-excitation BCDs with a lower
neutral gas column density and increasing turbulent motions. These two factors 
favour a more efficient escape of the He~{\sc i} Ly$\alpha$ $\lambda$584\AA\ 
photons which are at the origin of the He~{\sc i} $\lambda$2.058$\mu$m emission.

\section*{Acknowledgements}  
Support for this work was provided by NASA grants 
1463350 and GO4-15084X. 
Y.I.I. is grateful to the staff of the University of Virginia
for their warm hospitality. 
Funding for the SDSS and SDSS-II has been provided by the Alfred P. Sloan Foundation, the Participating Institutions, the National Science Foundation, the U.S. Department of Energy, the National Aeronautics and Space Administration, the Japanese Monbukagakusho, the Max Planck Society, and the Higher Education Funding Council for England. The SDSS Web Site is http://www.sdss.org/.
    The SDSS is managed by the Astrophysical Research Consortium for the Participating Institutions. 


\bsp

\label{lastpage}

\end{document}